\documentclass[12pt]{article}

\usepackage{cite}
\usepackage{axodraw}
\usepackage[dvips]{graphicx}
\usepackage{epsfig}

\def\theequation{\arabic{section}.\arabic{equation}}

\begin{document}

\begin{flushright}
CERN-PH-TH/2006-003\\[-2pt]
{\tt hep-ph/0601080}\\
January 2006
\end{flushright}
\bigskip

\begin{center}
{\LARGE {\bf {\boldmath $F_D$}-Term  Hybrid Inflation with}}\\[0.3cm] 
{\LARGE {\bf Electroweak-Scale Lepton Number Violation}}\\[1.5cm] 
{\large Bj\"orn Garbrecht$^{\, a}$ and Apostolos Pilaftsis$^{\, a,b}$}\\[0.5cm]
{\em $^a$School of Physics and Astronomy, University of Manchester,}\\ 
{\em Manchester M13 9PL, United Kingdom}\\[0.3cm]
{\em $^b$CERN, Physics Department, Theory Division, CH-1211 Geneva 23, 
Switzerland}
\end{center}

\vspace{1.5cm} \centerline{\bf ABSTRACT} 

\noindent 
We study $F$-term hybrid inflation in a novel supersymmetric extension
of the SM with a subdominant Fayet--Iliopoulos $D$-term.  We call this
particular form  of inflation, in short,  $F_D$-term hybrid inflation.
The  proposed  model  ties  the  $\mu$-parameter of  the  MSSM  to  an
SO(3)-symmetric  Majorana mass $m_N$,  through the  vacuum expectation
value  of the  inflaton  field.   The late  decays  of the  ultraheavy
particles  associated  with the  extra  U(1)  gauge  group, which  are
abundantly  produced  during the  preheating  epoch,  could lower  the
reheat temperature  even up to  1~TeV, thereby avoiding  the gravitino
overproduction problem.   The baryon asymmetry in the  Universe can be
explained by thermal electroweak-scale resonant leptogenesis, in a way
independent  of any pre-existing  lepton- or  baryon-number abundance.
Further   cosmological  and   particle-physics  implications   of  the
$F_D$-term hybrid model are briefly discussed.

\noindent


\newpage

\setcounter{equation}{0}
\section{Introduction}

The  inflationary   paradigm  constitutes  an   ingenious  theoretical
framework,  in which  many  of the  outstanding  problems in  standard
cosmology  can be  successfully  addressed~\cite{review}.  The  recent
WMAP    data~\cite{WMAP},    compiled    with    other    astronomical
observations~\cite{MT,Lyman}, improved  upon the precision  of about a
dozen of  cosmological parameters.   These include the  power spectrum
$P^{1/2}_{{\cal  R}}$ of curvature  perturbations, the  spectral index
$n_s$,  the baryon-to-photon  ratio of  number densities  $\eta_B$ and
others.   The  values of  these  cosmological  observables put  severe
constraints on  the model-building  of successful models  of inflation
and  their theoretical  parameters.  For  instance, one  of  the basic
requirements for  slow-roll inflation  is that the  so-called inflaton
potential be flat.  In this respect, supersymmetry (SUSY) emerges as a
compelling ingredient in model-building for protecting the flatness of
the inflaton potential against quantum corrections.

In addition to the aforementioned element of naturalness, inflationary
models would have a greater value if they were predictive and testable
as well.  One  such predictive and perhaps most  appealing scenario is
the well-celebrated  model of hybrid  inflation~\cite{Linde}.  In this
model, the inflaton  field $\phi$ can start its  slow-roll from values
well  below the  Planck scale  $m_{\rm Pl}  =  2.4\times 10^{18}$~GeV.
This renders the model very  predictive, in the sense that an infinite
set of possible higher-dimensional non-renormalizable operators, being
suppressed by  inverse powers of $1/m_{\rm Pl}$,  will not generically
contribute   significantly  to   cosmological  observables,   such  as
$P^{1/2}_{{\cal R}}$  and $n_s$.  In the hybrid  model, inflation ends
through  the  so-called waterfall  mechanism,  once  the field  $\phi$
passes below  a critical value  $\phi_c$.  When this  happens, another
field  $X$  different  from  $\phi$,  with  vanishing  initial  value,
develops  a tachyonic  instability and  rolls  fast down  to its  true
vacuum    expectation    value~(VEV).    Super-\linebreak    symmetric
realizations of  hybrid inflation  from $F$-terms were  first analyzed
in~\cite{CLLSW,DSS}, whereas hybrid  inflation triggered by a dominant
Fayet--Iliopoulos~(FI) $D$-term~\cite{FI}  was subsequently considered
in~\cite{Halyo}.

In  this  paper  we  study   $F$-term  hybrid  inflation  in  a  novel
supersymmetric extension  of the  Standard Model~(SM) that  includes a
subdominant  FI  $D$-term. We  call  this  scenario  for brevity,  the
$F_D$-term hybrid model.  To account for the low-energy neutrino data,
we introduce 3 singlet neutrino superfields $\widehat{N}_{1,2,3}$ that
contain   3   right-handed   neutrinos  $\nu_{1,2,3\,R}$   and   their
supersymmetric  scalar   counterparts  $\widetilde{N}_{1,2,3}$.   Most
importantly,  the  model  ties  the  $\mu$-parameter  of  the  Minimal
Supersymmetric  Standard Model~(MSSM) to  an SO(3)  symmetric Majorana
mass    $m_N$,   through    the    VEV   of    the   inflaton    field
$\phi$~\cite{PU2,Francesca}.   Hence,   the  $F_D$-term  hybrid  model
naturally predicts lepton-number  violation at the TeV or  even at the
electroweak scale. In this  scenario, the non-zero baryon asymmetry in
the  Universe (BAU),  $\eta_B  \approx 6.1  \times  10^{-10}$, can  be
explained  by  leptogenesis~\cite{FY,BAUpapers}  and  specifically  by
thermal electroweak-scale resonant leptogenesis~\cite{APRD,PU2}.

In this paper  we also study the constraints on  the parameters of the
$F_D$-term hybrid model that  result from a reheat temperature $T_{\rm
reh} \stackrel{<}{{}_\sim} 10^9$~GeV, which  is necessary to avoid the
well-known gravitino overproduction  problem.  This consideration puts
severe  limits on  the size  of  the superpotential  couplings of  the
theory, forcing  them {\em all}  to acquire rather  suppressed values,
namely to be smaller than about $10^{-5}$~\cite{SS}.  To overcome this
problem of unnaturalness, the presence of a subdominant FI $D$-term in
the  $F_D$-term  hybrid model  is  very  crucial  and provides  a  new
mechanism  of  relaxing  dramatically  the above  upper  limit.   More
explicitly, the size  of the $D$-term controls the  decay rates of the
ultraheavy fermions  and bosons associated with the  extra gauge group
U(1)$_X$.    In  the   absence   of  the   $D$-term   and  any   other
non-renormalizable   interaction,    these   ultraheavy   gauge-sector
particles are  absolutely stable. On  the other hand,  these particles
are abundantly  produced during the preheating  epoch, thus dominating
the energy  density of  the Universe shortly  after the period  of the
first   reheating  caused   by  the   perturbative   inflaton  decays.
Therefore, their late decays induced by a non-vanishing $D$-term could
give rise  to a second reheating  phase in the evolution  of the early
Universe. Depending on the actual size of the FI $D$-term, this second
reheat temperature may be as low  as 1~TeV, giving rise to an enormous
entropy  release that can  dilute the  gravitinos produced  during the
first reheating to an unobservable level.

The paper is organized  as follows: Section~\ref{FDmodel} presents the
model-building   aspects   of  the   $F_D$-term   hybrid  model   with
electroweak-scale  lepton-number violation. Technical  details related
to the  radiatively-induced FI  $D$-term are relegated  to Appendix~A.
In Section~\ref{reheat},  we estimate the reheat  temperature from the
perturbative  inflaton  decays  and  derive  the  resulting  gravitino
constraint  on  the  theoretical  parameters.   We  then  discuss  the
non-perturbative production of the supermassive fields associated with
the U(1)$_X$  gauge group  during the preheating  epoch and  how their
late decays can help to lower the reheat temperature even up to~1~TeV.
Section~\ref{inflation} is devoted  to inflation.  Here we investigate
two regimes: (i) cold  hybrid inflation, where dissipative effects can
be ignored, and (ii)  warm hybrid inflation, where dissipative effects
dominate   over   the   expansion    rate   of   the   Universe.    In
Section~\ref{BAU} we  illustrate how  the $F_D$-term hybrid  model can
realize  thermal  electroweak-scale  resonant leptogenesis.   Finally,
Section~\ref{conclusions}  summarizes  our  conclusions,  including  a
brief discussion  of further  possible implications of  the $F_D$-term
hybrid model for particle physics and cosmology.

\setcounter{equation}{0}
\section{The {\boldmath $F_D$}--Term Hybrid Model}\label{FDmodel}

The $F_D$-term hybrid model may be defined by the superpotential
\begin{eqnarray}
  \label{Wmodel}
 W & =& \kappa\, \widehat{S}\, \Big( \widehat{X}_1
\widehat{X}_2\:  -\: M^2\Big)\ +\ \lambda\, \widehat{S} \widehat{H}_u
\widehat{H}_d\ +\ \frac{\rho}{2}\, \widehat{S}\, \widehat{N}_i
\widehat{N}_i\ +\ h^{\nu}_{ij} \widehat{L}_i \widehat{H}_u
\widehat{N}_j\nonumber\\ &&+\ W_{\rm MSSM}^{(\mu = 0)}\; ,
\end{eqnarray} where $W_{\rm MSSM}^{(\mu = 0)}$ is the MSSM superpotential
without the $\mu$ term: 
\begin{equation} W_{\rm MSSM}^{(\mu = 0)}\ =\
  h^u_{ij}\,\widehat{Q}_i\widehat{H}_u\widehat{U}_j\: +\: 
h^d_{ij}\,\widehat{H}_d\widehat{Q}_i\widehat{D}_j\: +\:
  h_l\, \widehat{H}_d\widehat{L}_l\widehat{E}_l \; . 
\end{equation}
In~(\ref{Wmodel}),   $\widehat{S}$    is   the   SM-singlet   inflaton
superfield,  and  $\widehat{X}_1$  and  $\widehat{X}_2$  is  a  chiral
multiplet pair  of the so-called waterfall fields  which have opposite
charges   under   the    additional   U(1)$_X$   gauge   group.    The
superpotential~(\ref{Wmodel}) of  the model is  uniquely determined by
the $R$  transformation: $\widehat{S} \to  e^{i\alpha}\, \widehat{S}$,
$\widehat{X}_{1,2}     \to     e^{\pm     i\beta}\,\widehat{X}_{1,2}$,
$\widehat{L}   \to   e^{i\alpha}\,   \widehat{L}$,  $\widehat{Q}   \to
e^{i\alpha}\, \widehat{Q}$,  with $W  \to e^{i\alpha} W$,  whereas all
other  fields remain  invariant  under an  $R$  transformation.  As  a
consequence of  the $R$ symmetry, higher-dimensional  operators of the
form  $\widehat{X}_1 \widehat{X}_2  \widehat{N}_i \widehat{N}_i/m_{\rm
Pl}$, for example, are not allowed.

In   addition,  the   model  contains   a  subdominant   FI  $D$-term,
$-\frac{1}{2}  g\, m^2_{\rm  FI}\,  D$, giving  rise  to the  $D$-term
potential
\begin{equation}
  \label{Dterm}
V_D\ =\ \frac{g^2}{8}\ \Big( |X_1|^2\, -\, |X_2|^2\, -\, m^2_{\rm
  FI}\,\Big)^2\; .
\end{equation}
The FI $D$-term plays no role in the inflationary dynamics, as long as
$g m_{\rm  FI} \ll \kappa\, M$.  In  Appendix~\ref{Dappendix}, we show
how  a  subdominant  $D$-term   can  be  generated  radiatively  after
Planck-scale heavy  degrees of freedom  have been integrated  out. The
presence of the $D$-term is  important to break an accidental discrete
charge symmetry that survives  after the spontaneous symmetry breaking
(SSB)  of the  U(1)$_X$.  Such  a breaking  is crucial  to  render all
U(1)$_X$  gauge-sector   particles  unstable.   As  we   will  see  in
Section~\ref{reheat}, an  upper limit  on the size  of the FI  term is
obtained   by  requiring  a   sufficiently  low   reheat  temperature,
e.g.~$T_{\rm  reh}   \stackrel{<}{{}_\sim}  10^9$~GeV,  in   order  to
suppress the gravitino abundance to an unobservable level.

From~(\ref{Wmodel}) it  is straightforward  to read the  Lagrangian of
the inflationary soft SUSY-breaking sector,
\begin{equation}
  \label{Lsoft}
-\, {\cal L}_{\rm soft}\ =\ M^2_S S^*S\: +\: \Big( 
\kappa A_\kappa\, S X_1X_2\: +\: \lambda A_\lambda S H_u H_d\: \: +\:
\frac{\rho}{2}\, A_\rho\, S \widetilde{N}_i\widetilde{N}_i\: 
-\: \kappa a_S M^2 S \: \ +\ {\rm  H.c.}\,\Big)\,,
\end{equation}
where   $M_S$,   $A_{\kappa,\lambda,\rho}$    and   $a_S$   are   soft
SUSY-breaking mass parameters of  order $M_{\rm SUSY} \sim 1$~TeV.  

In  the  regime  $|S|  \gg  M$ relevant  to  inflation,  the  dominant
tree-level  and one-loop contributions  to the  renormalized effective
potential may be described by
\begin{eqnarray}
  \label{VpotFD}
V_{\rm inflation} & \approx & \kappa^2 M^4\ \Bigg[\, 1\: +\: 
\frac{1}{64\pi^2}\, \Big(\, 4\kappa^2 \: +\: 8\lambda^2\: +\:
6\rho^2\,\Big)\,  \ln\Bigg(\frac{|S|^2}{M^2}\Bigg)\,\Bigg]\nonumber\\
&&
-\: \Big( \kappa a_S M^2 S\: +\: {\rm H.c.}\Big)\ +\ V_{\rm SUGRA}\ ,
\end{eqnarray}
where $V_{\rm SUGRA}$ denotes the supergravity (SUGRA) correction that
results  from the  K\"ahler  potential.  Assuming  a minimal  K\"ahler
potential, the SUGRA correction of  interest to us takes on the simple
form~\cite{CLLSW,CP,LR}
\begin{equation}
  \label{Vsugra}
V_{\rm SUGRA}\ =\ \kappa^2 M^4\, \frac{|S|^4}{2\,m^4_{\rm Pl}}\ ,
\end{equation}
where $m_{\rm Pl} = 2.4\times 10^{18}$~GeV is the reduced Planck mass.
Possible   one-loop   contributions   to  $V_{\rm   inflation}$   from
$A_{\kappa,\lambda,\rho}$-terms become significant only for relatively
low  values  of  $M$,  e.g.   $M\stackrel{<}{{}_\sim}  10^8$~GeV,  for
$\kappa,\lambda,\rho \sim  1$, and may therefore be  neglected. At the
tree level,  however, only the tadpole  term $\kappa a_S  M^2\, S$ may
become relevant for  values of $\kappa \stackrel{<}{{}_\sim} 10^{-4}$,
whereas  the  other soft  SUSY-breaking  terms  are negligible  during
inflation~\cite{SS}.

We now investigate the stability of the inflationary trajectory in the
presence of  the Higgs doublets $H_{u,d}$ and  the right-handed scalar
neutrinos   $\widetilde{N}_{1,2,3}$.     Specifically,   the   initial
condition for inflation is
\begin{equation}
  \label{initial}
{\rm Re}\, S^{\rm in}\ =\ |S^{\rm in}|\ \gg\ M\,,\qquad 
X^{\rm in}_{1,2}\ =\ 0\,,\qquad
H^{\rm in}_{u,d}\ =\ 0\,,\qquad
\widetilde{N}^{\rm in}_{1,2,3}\ =\ 0\; . 
\end{equation} 
At the end  of inflation, one should ensure  that the waterfall fields
acquire a  high VEV, i.e. $X^{\rm  end}_{1,2}\ =\ M$,  while all other
fields have small VEVs, possibly  of the electroweak order. To achieve
this, we have to require that the Higgs-doublet and the sneutrino mass
matrices stay positive definite throughout the inflationary trajectory
up to the critical value  $|S_c| \approx M$, whereas the corresponding
mass  matrix of  $X_{1,2}$ will  be the  first to  develop  a negative
eigenvalue and tachyonic instability close to $|S_c|$. In this way, it
will be the fields $X_{1,2}$ which will first start moving away from 0
and set in  to the `good' vacuum $X^{\rm end}_1\  =\ X^{\rm end}_2\ =\
M$,   instead  of   having  the   other  fields,   e.g.~$H_{1,2}$  and
$\widetilde{N}^{\rm in}_{1,2,3}$,  go to a `bad'  vacuum where $X^{\rm
end}_{1,2}\ =\ 0$,  $H^{\rm end}_{1,2}\ =\ \frac{\kappa}{\lambda}\, M$
and  $\widetilde{N}^{\rm in}_{1,2,3}  = \frac{\kappa}{\rho}\,  M$.  To
see  this,  let  us write  down  the  mass  matrix in  the  background
Higgs-doublet field space $(H^\dagger_d\,,\ H_u )$ as
\begin{equation}
  \label{Mdoublet}
M^2_{\rm Higgs}\ =\ \left(\! \begin{array}{cc}
|\lambda|^2 |S|^2 & -\,\kappa \lambda (M^2 - X_1 X_2 )\: +\: \lambda
A_\lambda S \\
-\,\kappa^* \lambda^* (M^2 - X^*_1 X^*_2 ) +\: \lambda^*
A^*_\lambda S^* & |\lambda|^2 |S|^2 \end{array}\!\right)\ .
\end{equation}
The requirement  of positive definiteness  may be translated  into the
simple condition:
\begin{equation}
  \label{Scondition}
|\lambda|\, |S|^2\ \ge\ |\kappa (M^2 - X_1 X_2 )\: -\: 
A_\lambda S|\ .
\end{equation}
From  this last  inequality, we  may see  that the  condition $\lambda
\stackrel{>}{{}_\sim}   \kappa$  is   sufficient  for   ending  hybrid
inflation  to the `good'  vacuum.  Likewise,  one obtains  a condition
analogous to~(\ref{Scondition}) from  the sneutrino mass matrix, which
amounts to having $\rho  \stackrel{>}{{}_\sim} \kappa$.  The above two
restrictions on the superpotential couplings $\lambda$ and $\rho$ will
be imposed throughout our analysis.

As  mentioned above,  after  the  end of  inflation,  one has  $X^{\rm
end}_{1,2} =  M$, giving rise to  a high mass for  the inflaton field,
i.e.~$2|\kappa  |^2 M^2  |S|^2$.  Combining  this fact  with  the soft
SUSY-breaking tadpole  $-\kappa a_S M^2 S$ and  the trilinear coupling
$\kappa A_\kappa  S X^{\rm end}_1 X^{\rm  end}_2$, one gets  a VEV for
the inflaton~\cite{DLS}:
\begin{equation}
  \label{Send}
v_S\  \equiv\  \langle S^{\rm  end}  \rangle\  =\  \frac{1}{2|\kappa|}\,
\Big|\,A_{\kappa} - a_S\,\Big|\ ,
\end{equation}
where we have neglected the VEVs of the Higgs doublets $H_{u,d}$.  The
VEV  of $S$  induces an  effective $\mu$-term  and an  SO(3) symmetric
lepton-number-violating  Majorana   mass  $m_N$  of   the  electroweak
order~\cite{PU2}:
\begin{equation}
  \label{mumN}
\mu\ =\ \lambda\, v_S\;, \qquad m_N\ =\ \rho\, v_S\; .
\end{equation}
If $\rho$  and $\lambda$ are  comparable in magnitude, then  these two
mass  parameters  are  tied  together  and can  naturally  be  of  the
TeV or even of the electroweak scale.

In  Sections~\ref{reheat}  and~\ref{inflation},  we  will  derive  the
constraints  on the  key theoretical  parameters  $\kappa$, $\lambda$,
$\rho$  and $M$  from the  requirement  of a  low reheat  temperature,
$T_{\rm   reh}   \stackrel{<}{{}_\sim}   10^9$~GeV,   and   successful
inflation.

\setcounter{equation}{0}
\section{Preheating and Second Reheating}\label{reheat}

In the SUGRA  framework, the reheat temperature is  constrained by the
fact that  an overabundant amount  of gravitinos may  destroy, through
their  possible late decays,  the successful  predictions of  Big Bang
nucleosynthesis~\cite{Sarkar}.   This possibility  is avoided,  if the
gravitino  abundance  $Y_{3/2}$   is  small  enough,  i.e.~$Y_{3/2}  <
10^{-12}$.   The latter may  be translated  to an  upper limit  on the
reheat temperature, i.e.~$T_{\rm reh} \stackrel{<}{{}_\sim} 10^9$~GeV.
If the  gravitinos are stable, the  above limit may be  relaxed by one
order of magnitude to $\sim 10^{10}$~GeV.  This depends on whether the
so-called next-to-lightest supersymmetric  particle (NLSP) has a small
branching fraction to hadronic decay modes~\cite{FIY}.  In addition to
the above  upper limit, the  reheat temperature $T_{\rm reh}$  is also
constrained from  below, depending  on the mechanism  of baryogenesis.
Thus, for successful  electroweak-scale resonant leptogenesis, a lower
bound of order TeV on $T_{\rm reh}$ should be considered.

In the following we will study the post-inflationary dynamics. To this
end, let us define the fields:
\begin{eqnarray}
  \label{Xpm}
X_\pm \!&=&\! \frac{1}{\sqrt{2}}\, (X_1\: \pm\: X_2)\ =\  
\langle X_\pm \rangle\: +\: \delta X_\pm\; ,\nonumber\\
\delta X_\pm \!&=&\! \frac{1}{\sqrt{2}}\, 
(R_\pm\, +\, {\rm i}I_\pm)\; .
\end{eqnarray}
As mentioned  in the introduction,  inflation ends, once  the inflaton
field, $\phi = \sqrt{2}\, {\rm  Re}\, S$, rolls below a critical value
$\phi_c  \approx \sqrt{2}\,  M$.  Then,  the waterfall  regime begins,
where the  waterfall fields $S$ and  $R_+$ evolve rapidly  (we use the
gauge  freedom to  ensure that  all  VEVs point  to real  directions).
Ignoring small corrections due to a non-vanishing FI $D$-term, $m_{\rm
FI}$, the VEVs of $S$  and $R_+$ oscillate around \emph{zero}, whereas
$X_+$  attains  its  U(1)$_X$-breaking  VEV, $\langle  X_+  \rangle  =
\sqrt{2} M$.

The  masses of  the waterfall-  or $\kappa$-sector  fields  $\phi$ and
$R_+$ are equal to $m_\kappa = \sqrt 2 \kappa M$.  The inflaton $\phi$
decays  predominantly into  pairs  of charged  and neutral  higgsinos,
$\tilde{h}^\pm_{u,d}$, $\tilde{h}^0_{u,d}$, $\tilde{\bar{h}}^0_{u,d}$,
and     into    pairs     of    right-handed     Majorana    neutrinos
$\nu_{1,2,3\,R}$. The decay width of the inflaton is given by
\begin{equation}
  \label{infldecay}
\Gamma_\phi\ =\ \frac{1}{32\pi}\:  \Big(\, 4\lambda^2\: +\: 3 \rho^2\,
\Big)\: m_\kappa\; .
\end{equation}
It turns out that the field $R_+$ decays into the scalar SUSY partners
of the aforementioned fields at the same rate. Hence, we find
\begin{equation}
\Gamma_\phi\ =\ \Gamma_{R_+}\ \equiv\ \Gamma_\kappa\; .
\end{equation}
The reheat  temperature resulting from the perturbative  decays of the
$\kappa$-sector fields may usually be estimated by
\begin{equation}
  \label{Treh}
T^\kappa_{\rm reh}\ =\ \left( \frac{90}{\pi^2\, g_*}\right)^{1/4}\,
\sqrt{\Gamma_\kappa\: m_{\rm Pl} }\ ,
\end{equation}
where  $g_* = 228.75$  is the  number of  the relativistic  degrees of
freedom in the supersymmetric model under consideration. The gravitino
bound then implies that
\begin{equation}
  \label{Tkappa}
\kappa\, \left(\, \lambda^2\: +\: \frac{3}{4}\, \rho^2\, \right)\  
\stackrel{<}{{}_\sim}\ 3 \cdot 10^{-15}\,\times\, 
\left(\frac{T^\kappa_{\rm reh}}{10^9~{\rm GeV}}\right)^2\,
\left( \frac{10^{16}~{\rm GeV}}{M}\right)\; .
\end{equation}
If $\kappa \approx  \lambda \approx \rho$, this amounts  to being each
individual coupling smaller than about $10^{-5}$, for $M =10^{16}$~GeV
and $T^\kappa_{\rm reh} \stackrel{<}{{}_\sim} 10^9$~GeV.

So far, we have only  considered the post-inflationary dynamics of the
$\kappa$-sector fields, $S$, $R_+$ and  $I_+$, to which all the energy
of the inflationary potential is  stored at the onset of the waterfall
regime.  We  now turn our attention  to the $g$-sector,  namely to the
particles  associated  with  the  extra U(1)$_X$  gauge  group.   This
distinction of  the different fields involved after  inflation is made
clear  in Table~\ref{spectrum}.   Thus,  the $g$-  or U(1)$_X$  gauge-
sector contains  the U(1)$_X$ gauge  boson $V_\mu$, the  Dirac fermion
$\psi_g$, which  consists of the  gaugino $\lambda$ and  the fermionic
superpartner  of $X_-$,  and the  scalars $R_-$  and $I_-$;  the field
$I_-$  is  a  massless  would-be  Goldstone boson  which  becomes  the
longitudinal component  of $V_\mu$.  Each of  the $g$-sector particles
has a mass $m_g=2^{-1/2} g  \langle X_+ \rangle$.  In fact, during the
waterfall transition, their masses evolve rapidly from 0 to $g M$.  As
we will  see below, this  rapid non-adiabatic mass  variation triggers
the  so-called  preheating  mechanism,  through which  the  $g$-sector
particles can be produced in  sizeable amounts.  Their decays can only
be induced by  the presence of a non-vanishing  $D$-term, which breaks
explicitly a  discrete charge  symmetry in the  $F$- and  the $D$-term
sectors which would remain otherwise  intact even after the SSB of the
U(1)$_X$.

\begin{table}
\begin{center}
\begin{tabular}{|c|c|c|c|}
\hline
 & & & \\
Sector & Boson & Fermion & Mass\\
 & & & \\
\hline\hline
 & & & \\
Waterfall &
$S$, $R_+$, $I_+$ &
$\psi_\kappa=
\left(
\begin{array}{c}
\frac{1}{\sqrt2}
\left[
\left(1-\frac{v}{2M}\right)\psi_{X_1}
+\left(1+\frac{v}{2M}\right)\psi_{X_2}
\right]
\\
\psi_S^\dagger
\end{array}
\right)
$
&
$\sqrt 2 \kappa M$
\\
($\kappa$-sector) & & & \\
 & & & \\
\hline
 & & & \\
U(1)$_X$ Gauge &
$V_\mu$, $R_-$ &
$\psi_g=
\left(
\begin{array}{c}
\frac{1}{\sqrt2}
\left[
\left(1+\frac{v}{2M}\right)\psi_{X_1}
-\left(1-\frac{v}{2M}\right)\psi_{X_2}
\right]
\\
-{\rm i}\lambda^\dagger
\end{array}
\right)
$
&
$g M$
\\
($g$-sector) & & & \\
 & & & \\
\hline
\end{tabular}
\end{center}
\caption{\em  Particle spectrum  of the  waterfall  and U(1)$_X$ gauge
sectors  after inflation,  where  $V_\mu$ denotes  the U(1)$_X$  gauge
boson and $\lambda$~its associate gaugino.}\label{spectrum}
\end{table}

To make this last point explicit, let us express the relevant $F$- and
$D$-term   potential  in   terms   of  the   fields  $X_\pm$   defined
in~(\ref{Xpm}):
\begin{equation}
  \label{VFD}
V_{FD}\ =\ \frac{\kappa^2}{4}\, \Big|\,X^2_+\: -\: X^2_-\:
-\: 2\,M^2\,\Big|^2\ +\ \frac{g^2}{8}\, \Big(\,
X^*_+ X_-\: +\: X^*_- X_+\: -\: m^2_{\rm FI}\, \Big)^2\; . 
\end{equation}
It  is obvious  that the  potential $V_{FD}$  possesses  an additional
discrete  charge symmetry  under  the transformation,  $X_\pm \to  \pm
X_\pm$, if  the FI  mass term  vanishes, $m^2_{\rm FI}  = 0$.   In the
absence of a FI term, this  symmetry will still survive even after the
SSB of  the U(1)$_X$  along the flat  direction $\langle  X_1\rangle =
\langle X_2 \rangle =  M$,\footnote{Observe that an analogous discrete
charge  symmetry also  survives after  SSB in  the  so-called $D$-term
inflationary model~\cite{Halyo},  where $M =  0$ and $m_{\rm  FI} \neq
0$.   In  this  case,  the  waterfall fields  $X_{1,2}$  transform  as
$X_{1,2}  \to  \pm X_{1,2}$,  while  their  VEVs  after inflation  are
$\langle X_1 \rangle = m_{\rm FI}$ and $\langle X_2 \rangle = 0$.}  or
equivalently  when $\langle  X_+ \rangle  = \sqrt{2}  M$  and $\langle
X_-\rangle = 0$.  As a  consequence, the U(1)$_X$ gauge boson $V_\mu$,
the scalar field $R_- =  \sqrt{2}\,{\rm Re} (X_-)$ and their fermionic
superpartner $\psi_g$ are all stable  with a mass $g M$.  This feature
is highly unsatisfactory for the hybrid model without a FI term, since
these particles can be produced in large numbers during the preheating
process, and since  they are very massive, they  could dominate and so
overclose the Universe at later times.

The presence of  the FI term $m_{\rm FI}$  breaks explicitly the above
discrete charge  symmetry and  so provides a  new decay  mechanism for
making these  particles unstable.  To  leading order in  the expansion
parameter $m_{\rm FI}/M$,  the potential $V_{FD}$ given in~(\ref{VFD})
can be minimized using the linear field decompositions
\begin{equation}
  \label{Xdec}
X_+\ =\ \sqrt{2}\,M\: +\: \delta X_+\,,\qquad 
X_-\ =\ \frac{v}{\sqrt{2}}\: +\: \delta X_-\ ,
\end{equation}
where  $v  =  m^2_{\rm  FI}/(2M)$. Table~\ref{spectrum}  exhibits  the
particle  spectrum of  the  waterfall and  U(1)$_X$  gauge sectors  to
leading order  in $m_{\rm FI}/M$.  Unlike  the case of  a vanishing FI
$D$-term,  the scalar field  $R_-$ of  mass $gM$  will now  decay into
pairs   of   two  lighter   scalars,   $R_+$   and   $I_+$,  of   mass
$\sqrt{2}\,\kappa M$,  assuming that $g  \gg \kappa$. The  strength of
this coupling is given by the Lagrangian
\begin{equation}
  \label{Lint}
{\cal L}_{\rm int}\  =\ \frac{g^2 m^2_{\rm FI}}{8 M}\;  R_-\, (R_+^2 +
I_+^2)\; .
\end{equation}
The $D$-term  induced decay  width of the  $R_-$ particle  can readily
be found to be
\begin{equation}
  \label{GammaR}
\Gamma_{R_-}\ =\ \frac{g^3}{128 \pi}\, \frac{m^4_{\rm FI}}{M^3}\ ,
\end{equation}
and  the  same  rate also  holds  true  for  the  decay of  $I_-$,  or
equivalently   for   the   longitudinal   polarization   of   $V_\mu$.
Correspondingly, the  decays of  the $g$-sector fermions  $\psi_g$ are
induced by the Lagrangian
\begin{equation}
{\cal L}_{\rm int}\ =\ -\, \frac{g}{8}\ \left(\frac{m_{\rm FI}}{M}\right)^2\,
\left(R_+ -{\rm i}I_+\right)\, 
\bar\psi_g\, \frac{1-\gamma_5}{2}\, \psi_\kappa\ +\ {\rm H.c.}
\end{equation}
Neglecting  soft SUSY-breaking, we  find that  $\Gamma_{\psi_g} =
\Gamma_{R_-} \equiv \Gamma_g$.

If  the   decay  rate  $\Gamma_g$  of  the   $g$-sector  particles  is
sufficiently low, they may dominate the energy density of the Universe
at later times, eventually leading  to a second reheating phase due to
their  out  of equilibrium  decays.   To  offer  an initial  estimate,
consider  that,   after  the  first  reheating,   the  energy  density
$\varrho_\kappa$ of the waterfall-sector fields gets distributed among
their decay products and so diluted as relativistic radiation $\propto
a^{-4}$, where  $a$ is the usual cosmological  scale factor describing
the  expansion  of  the   Universe.   Meanwhile,  the  energy  density
$\varrho_g$  of  the  ultraheavy  $g$-sector  particles  produced  via
preheating     scales    as     $\propto     a^{-3}$,    such     that
$\varrho_g/\varrho_\kappa\propto     a$.     Moreover,     during    a
radiation-dominated epoch, the dependence of the Hubble expansion rate
$H$ on $a$ is $H\propto a^{-2}$.  Let us therefore denote with $H_{\rm
reh}$  the Hubble  rate at  the first  reheating of  the  Universe and
$H_{\rm    eq}$    the    Hubble    rate    at    the    time,    when
$\varrho_g=\varrho_\kappa$.  Then, the U(1)$_X$ gauge-sector particles
will dominate the energy density of the Universe, when
\begin{equation}
  \label{condition:domination}
H_{\rm            eq}\           =\            H_{\rm           reh}\,
\left(\frac{\varrho^0_g}{\varrho^0_\kappa}\right)^2\ \gg\ \Gamma_g\;,
\end{equation}
where the  superscript $0$ stands  for the energy density  right after
preheating.  Note  that $\varrho_g/\varrho_\kappa$ is  conserved until
the  time   of  the  first  reheating,  since   both  $\varrho_g$  and
$\varrho_\kappa$ scale as $a^{-3}$ during this period.

\begin{figure}[t]
\begin{center}
\epsfig{file=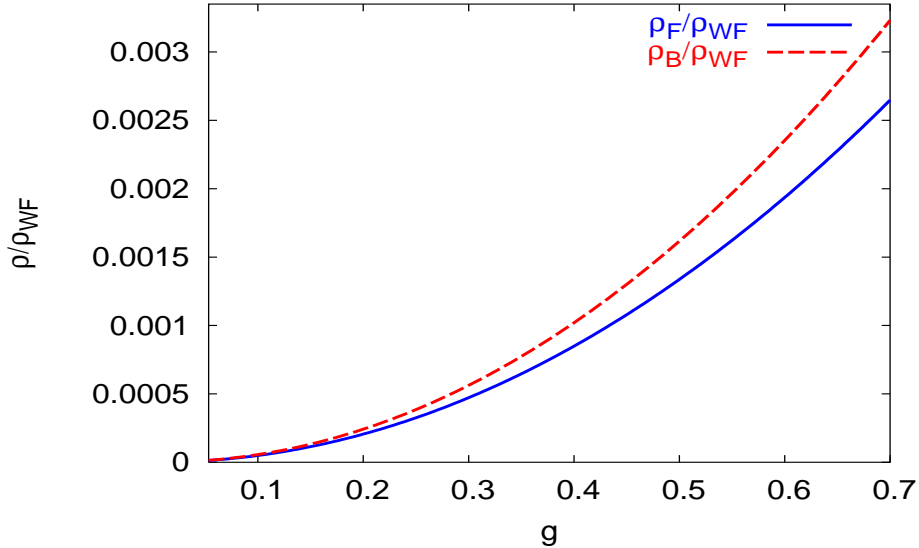, height=3.in,width=5in}
\end{center}
\caption{\em The  ratio of the  energy density $\varrho_F~(\varrho_B)$
of the  gauge-sector fermions~(bosons) produced via  preheating to the
energy density $\varrho_{\rm WF} \equiv \varrho_\kappa$ carried by the
waterfall-sector  particles  as  a  function  of $g$,  for  $\kappa  =
10^{-3}$.}\label{figure:preheating}
\end{figure}

The  $g$-sector particle  production  via preheating  can be  computed
numerically~\cite{PREHEATING}, by first solving for the mode functions
and then using these to calculate the Hamiltonian energy density.  For
the evolution  of the  VEVs $\langle X_1  \rangle \approx  \langle X_2
\rangle$, we assume that they  initially undergo strong damping due to
tachyonic   preheating~\cite{TACHYPREH}.   This   phenomenon   can  be
mimicked by setting
\begin{equation}
  \label{profile}
\langle X_1 \rangle\ =\ \langle X_2 \rangle\
=\ \left\{
\begin{array}{lr}
0\,, & \textnormal{for}\quad t\leq -\pi/(4 \sqrt 2 \kappa M)\,,\\
\frac{1}2\,M\, [1+\sin(\sqrt 2\kappa M t)]\,, & \textnormal{for}\quad
-\pi/(4 \sqrt 2 \kappa M)< t < \pi/(4 \sqrt2 \kappa M)\,,\\
M\, & \textnormal{for}\quad t\geq \pi/(4 \sqrt2 \kappa M)\,,
\end{array}
\right.
\end{equation}
More precise forms of field evolutions may be obtained using numerical
simulations~\cite{TACHYPREH}.  For an  initial estimate, however, only
the    velocity    of     the    transition    is    important.     In
Fig.~\ref{figure:preheating}   we   display   the   energy   densities
$\varrho_F$ and  $\varrho_B$ of  the $g$-sector fermions  $\psi_g$ and
bosons $R_-$ and $V_\mu$  (produced via preheating), normalized to the
energy density $\varrho_{\rm WF} \equiv \varrho_\kappa$ carried by the
waterfall-sector  particles,  as   functions  of  the  U(1)$_X$  gauge
coupling   $g$,    for   $\kappa   =   10^{-3}$.     For   the   given
profile~(\ref{profile}) of  field evolutions, these  normalized energy
densities depend only very weakly on $\kappa$.

The  above results  strongly  suggest that  the U(1)$_X$  gauge-sector
particles,  $\psi_g$, $R_-$ and  $V_\mu$, if  sufficiently long-lived,
will  dominate the  energy  density  of the  early  Universe.  We  may
estimate the second reheat temperature $T^g_{\rm reh}$ caused by their
late decays,  by employing a  formula very analogous  to (\ref{Treh}).
Solving this last relation for the ratio $m_{\rm FI}/M$ yields
\begin{equation}
  \label{ratio:mFI:M}
\frac{m_{\rm FI}}{M} \approx \ 8.4 \cdot 10^{-4}\times
\left( \frac{0.5}{g}\right)^{3/4}
\left(\frac{T^g_{\rm reh}}{10^9~{\rm GeV}}\right)^{1/2}\,
\left( \frac{10^{16}~{\rm GeV}}{M}\right)^{1/4}\; .
\end{equation}
For second reheat  temperatures of cosmological interest, i.e.~$1~{\rm
TeV}\leq  T^g_{\rm reh}\leq  10^9~{\rm GeV}$,  we obtain  the combined
constraint for $M = 10^{16}$~GeV:
\begin{equation}
  \label{FIcombined}
10^{-6}\ \stackrel{<}{{}_\sim}\ \frac{m_{\rm FI}}{M}\
\stackrel{<}{{}_\sim}\ 10^{-3}\; .
\end{equation}
From  our discussion  in this  section, it  is evident  that  the late
decays  of  the ultraheavy  U(1)$_X$  gauge-sector  fields, which  are
copiously produced  during the preheating  epoch, will give rise  to a
second reheating  phase in  the evolution of  the early Universe  at a
temperature $T^g_{\rm  reh} \ll  T^\kappa_{\rm reh}$.  This  makes the
$F_D$-term  hybrid  model an  interesting  cosmological scenario  that
could   even   lead  to   a   complete   relaxation   of  the   strict
bound~(\ref{Tkappa}) on the  couplings $\kappa,\ \lambda,\ \rho$.  The
reason is that gravitinos, which are produced very efficiently at high
reheat temperatures  $T^\kappa_{\rm reh}>10^9 {\rm GeV}$,  will now be
diluted  by the  large entropy  release from  the late  decays  of the
$g$-sector   particles.   In   this  way,   the   so-called  gravitino
overproduction problem can be completely avoided.  A~detailed study of
this topic will be given elsewhere~\cite{GPP}.

\setcounter{equation}{0}
\section{Inflation}\label{inflation}

In  this section  we will  discuss the  additional constraints  on the
theoretical parameters  of the $F_D$-term hybrid model  from the power
spectrum  $P_{\cal  R}^{1/2}$  and   the  spectral  index  $n_s$.   We
distinguish  two possible regimes  of inflation:  (i) the  cold hybrid
inflation   (CHI),  where   dissipative  effects   on   inflation  are
negligible, e.g.~for $\kappa ,\ \lambda ,\ \rho\ \stackrel{<}{{}_\sim}
10^{-2}$ and  (ii) the warm hybrid inflation  (WHI), where dissipation
might dominate over the expansion rate of the Universe~\cite{AB}.

\subsection{Cold Hybrid Inflation}

In models of hybrid inflation, the spectral index $n_s$ may well be
approximated as follows~\cite{review}:
\begin{equation}
  \label{nS}
n_s\: -\: 1\ =\ \frac{d\ln P^{1/2}_{\cal R}}{d\ln k}\ \approx\ 2\eta\; , 
\end{equation}
where $k$ is the comoving wavenumber at the horizon exit and
\begin{equation}
  \label{eta}
\eta\ =\ m^2_{\rm Pl}\ \frac{V_{\phi\phi}}{V}\ 
\end{equation} 
is  the so-called  $\eta$-parameter. In~(\ref{eta}),  $V$  denotes the
inflationary  potential,  and  $V_\phi  = dV/d\phi$,  $V_{\phi\phi}  =
d^2V/d\phi^2$  etc. The  current WMAP  data~\cite{WMAP} show  a strong
preference for a  red-tilted spectrum, with $n_s -  1 \le 0$, implying
that  $V_{\phi\phi}  \le 0$.  The  actual value  is  $n_s  = 0.98  \pm
0.02$~\cite{Lyman}.

The  $T_{\rm   reh}$  constraint~(\ref{Tkappa})  on   the  theoretical
parameters imply  that $\kappa,\ \lambda,\  \rho \stackrel{<}{{}_\sim}
10^{-5}$.   In this case,  the radiative  correction to  the potential
becomes subdominant and  may be ignored to a  good approximation.  The
potential driving inflation simplifies considerably to
\begin{equation}
  \label{VCHI}
V_{\rm inflation}\ =\ \kappa^2 M^4\: -\: \sqrt{2}\,\kappa\,a_S M^2 \phi\:
+\: \frac{1}{2}\, M^2_S\, \phi^2\: +\:  \frac{\kappa^2\,M^4}{8\,m^4_{\rm
    Pl}}\, \phi^4\ ,
\end{equation}
where  $\phi  =  \sqrt{2}\,  {\rm  Re}\,  S$  is  the  inflaton  field
canonically normalized. For $M_S < 1$~TeV, $\kappa \ge 10^{-6}$ and $M
\ge 10^{15}$~GeV,  the soft SUSY-breaking  term $M_S$ can  be omitted.
The  inflationary   potential  $V_{\rm  inflation}$   of  (\ref{VCHI})
generically leads to a blue-tilted spectrum,  i.e.~$n_s - 1 = 2 \eta >
0$, which is slightly disfavoured by the recent WMAP data.

In the  following, we  will concentrate on  the regime where  the loop
correction dominates the slope of  the potential, such that a negative
value for  $n_s -  1$ becomes possible.   This possibility  arises for
$10^{-4}   \stackrel{<}{{}_\sim}\   \kappa   ,\   \lambda   ,\   \rho\
\stackrel{<}{{}_\sim}  10^{-2}$.  Naively,  such large  values  of the
parameters  lead  to a  too  high  reheat  temperature $T_{\rm  reh}$,
i.e.~$T_{\rm reh} \stackrel{>}{{}_\sim}  10^{10}$~GeV.  However, as we
have discussed in Section~\ref{reheat},  the presence of a subdominant
$D$-term renders the stable U(1)$_X$ gauge-sector fields unstable, and
so a large  amount of entropy can be released  from their late decays,
leading to a $T_{\rm reh}$ which may even be as low as 1~TeV.

Our results simplify considerably if one assumes that the slope of the
inflationary  potential given  in~(\ref{VpotFD}) is  dominated  by the
$\lambda$-dependent  term. To  be  specific, the  number of  $e$-folds
${\cal N}_e$ is given by
\begin{equation}
  \label{Nefold}
{\cal N}_e\ =\ \frac{1}{m^2_{\rm Pl}}\; \int_{\phi_{\rm
    end}}^{\phi_{\cal N}}\, d\phi\: \frac{V}{V_\phi}\ \approx\ 
\frac{2\pi^2}{\lambda^2}\; \frac{\phi^2_{\cal N}}{m^2_{\rm Pl}}\ .
\end{equation}
Notice that  at the horizon exit,  it is $\phi_{\cal  N} = \sqrt{{\cal
N}_e/2}\,   (\lambda/\pi)\,    m_{\rm   Pl}$   and    $\phi_{\cal   N}
\stackrel{<}{{}_\sim}    10^{-1}\,   m_{\rm    Pl}$,    for   $\lambda
\stackrel{<}{{}_\sim}  0.1$ and  ${\cal  N}_e =  60$. Hence  inflation
starts at values of $\phi_{\cal N}$ well below $m_{\rm Pl}$.  In terms
of  the   number  of  $e$-folds  ${\cal  N}_e$,   the  power  spectrum
$P^{1/2}_{\cal R}$ of the curvature perturbations may now be given by
\begin{equation}
  \label{PRCHI}
P^{1/2}_{\cal R}\ =\ \frac{1}{2\sqrt{3}\, \pi m^3_{\rm Pl}}\; 
\frac{V^{3/2}}{|V_\phi|}
\ \approx\ \sqrt{\frac{2{\cal N}_e}{3}}\
\frac{\kappa}{\lambda}\ \Bigg( \frac{M}{m_{\rm Pl}}\Bigg)^2\ =\
5\times 10^{-5}\ .
\end{equation}
Evidently, for  ${\cal N}_e =  60$ and $M=10^{16}$~GeV,  the parameter
$\lambda$ cannot  be by more than  one order of  magnitude larger than
$\kappa$, i.e.~$\lambda  \stackrel{<}{{}_\sim} 10\, \kappa$.  Finally,
the spectral index $n_s$ in terms  of ${\cal N}_e$ may be expressed as
follows:
\begin{equation}
  \label{etaCHI}
n_s\, -\, 1\ =\ -\ \frac{1}{{\cal N}_e}\ \approx\ -\,0.02\ ,
\end{equation}
for ${\cal N}_e  = 50$--60.  In this CHI regime,  the model predicts a
red-tilted spectrum, as currently favoured by the WMAP data.

\subsection{Warm Hybrid Inflation}

It has been extensively  argued~\cite{AB} that dissipative effects due
to  radiation production  of massless  particles during  inflation may
dominate over  the expansion  rate $H$ of  the Universe. This  form of
inflation is known as warm inflation. Although a firm first principles
derivation  for  the  existence  of  a strong  dissipative  regime  of
inflation is still missing,\footnote{A detailed calculation based on a
two-particle  irreducible effective  action in  an  expanding deSitter
background  metric would  be highly  preferable.}  it  might  be worth
presenting tentative results for  such a possible situation, using the
semi-empiric formalism on warm inflation developed in~\cite{AB}.

In  the  framework  of  WHI,  dissipation occurs  from  the  radiation
produced by the  decays of the excited $H_u$  doublet of mass $\lambda
S$.  Specifically, the interactions relevant to WHI are
\begin{equation}
  \label{Lwarm}
-\, {\cal L}_{\rm int}^{\rm WHI}\ =\ |S|^2\, \bigg[\, |\lambda |^2\,
  |H_u|^2\: +\: |\rho |^2\, \bigg(\,\sum_{i=1}^3\,
  |\widetilde{N}_i|^2\bigg)\,\bigg]\ +\ \Big( h_t\, H_u\, \bar{Q}_t\,
t_R\: +\: h^{\nu}_{ij}\, \bar{L}_i \tilde{h}_u
\widetilde{N}_j\: +\: {\rm H.c.}\Big)\; .
\end{equation}
The dominant decay mode will be $H_u \to Q_t t_R$~\cite{BB}; the other
possible  decay  channel  $\widetilde{N}_j  \to  L_i  \tilde{h}_u$  is
Yukawa-coupling suppressed and kinematically allowed only when $\rho >
\lambda$.   Adapting the  results  of~\cite{AB,BB} to  our model,  the
dominant friction term for $|S| \gg M$ is given by
\begin{equation}
  \label{Ys}
Y_S\ \approx\ \frac{\sqrt{\pi}\, \alpha_\lambda^{3/2}\,
\alpha_t}{20\,\sqrt{2}}\  \phi\; ,
\end{equation}
where   $\alpha_\lambda   =    \lambda^2/(4\pi)$   and   $\alpha_t   =
h^2_t/(4\pi)$.   The dynamics  of warm  inflation is  governed  by the
following two equations:
\begin{eqnarray}
  \label{phiS}
\ddot{\phi}\ +\ 3H\, ( 1 + r )\, 
\dot{\phi}\ +\ V_\phi & = & 0\; ,\\
  \label{rhorad}
\dot{\rho}_{\rm rad}\ +\ 4\, H\rho_{\rm rad} & = & Y_S\, \dot{\phi}^2\ ,
\end{eqnarray}
where  $r =  Y_S/(3H)$,  with $H^2  \approx  \kappa^2 M^4/(3  m^2_{\rm
Pl})$.  In  the strong dissipative  regime where $r \gg  1$, inflation
usually ends  when $\rho_{\rm rad}  > \rho_{\rm vac}  \approx \kappa^2
M^4$.

Assuming conditions of  slow roll during WHI, i.e.~$\eta  /r^2 \ll 1$,
we may determine the number of $e$-folds by 
\begin{equation}
  \label{NeWHI}
{\cal N}_e \ =\ \frac{1}{m^2_{\rm Pl}}\; \int^{\phi_{\cal
      N}}_{\phi_{\rm end}}\, d\phi\ \frac{(1+r)\,V }{V_\phi}\ =\
\frac{\pi \alpha_\lambda^{1/2}\, \alpha_t}{60\, \kappa}\ \frac{
      \phi^3_{\cal N}}{ m_{\rm Pl}\, M^2}\ .
\end{equation}
In the  limit $r\gg 1$, the  power spectrum $P_{\cal  R}^{1/2}$ due to
WHI is approximately given by
\begin{equation}
  \label{PRWHI}
(P_{\cal R}^{1/2})_{\rm WHI}\ \approx\ \left( \frac{3\pi}{4} \right)^{1/4}\,
\sqrt{\frac{T_{\rm rad}}{H}}\;
r^{5/4}\, (P_{\cal R}^{1/2})_{\rm CHI}\ .
\end{equation}
The  temperature $T_{\rm rad}$ associated with radiation production  can be
calculated from (\ref{rhorad}), by solving the approximate equation
\begin{equation}
  \label{Trad}
\rho_{\rm rad}\ =\ \frac{\pi^2}{30}\ g_*\ T^4_{\rm rad}\ \approx\
\frac{3r}{4}\ \dot{\phi}^2\; ,
\end{equation}
where $\dot{\phi} \approx - V_\phi/(3r H)$ is evaluated at the horizon
exit.  Putting everything together, we find
\begin{equation}
  \label{PRWHI2}
(P_{\cal R}^{1/2})_{\rm WHI}\ \approx\ g_*^{-1/8}\, {\cal N}_e^{5/8}\,
  (2\kappa)^{1/4}\, \alpha_\lambda^{5/8}\, \alpha_t^{1/2}\, 
  \left(\frac{M}{m_{\rm Pl}}\right)^{1/2}\ =\ 5 \times 10^{-5}\ .
\end{equation}
It is interesting  to observe that WHI leads  to a viable inflationary
scenario   even   for    strong   couplings,   e.g.~for   $\kappa   ,\
\alpha_\lambda,\    \alpha_t\   \sim~1$.     In    this   case,    the
U(1)$_X$-breaking  scale  $M$ will  be  as  low  as $10^{10}$~GeV,  in
agreement  with the  earlier  discussion in  \cite{BB}. Obviously,  it
would be  difficult to associate such  a low scale for  $M$ with gauge
coupling  unification. Finally,  the spectral  index $n_s$  in  WHI is
calculated in  terms of ${\cal  N}_e$ to be:  $n_s - 1 \approx  - 5/(4
{\cal N}_e) \approx -0.025$.

\setcounter{equation}{0}
\section{Baryon Asymmetry in the Universe}\label{BAU}

As discussed in Section~\ref{reheat},  the late decays of the U(1)$_X$
gauge-sector particles  may lead  to a second  reheating phase  in the
evolution  of the  early Universe,  giving rise  to a  very  low final
reheat  temperature~$T_{\rm  reh}$.   Depending  on the  size  of  the
$D$-term, $T_{\rm reh}$ may even be  as low as 1~TeV.  In such a case,
the  BAU  may  be  explained  by  thermal  electroweak-scale  resonant
leptogenesis~\cite{APRD,PU2}.  The $F_D$-term hybrid model under study
can realize  such a  scenario even within  a minimal  SUGRA framework,
where  all  soft  SUSY-breaking  parameters  are  constrained  at  the
gauge-coupling unification point $M_X$, which can be chosen to be $M =
M_X  \approx 10^{16}$~GeV.   Instead, electroweak  baryogenesis  is no
longer viable in minimal  SUGRA, since it requires an unconventionally
large   hierarchy  between  the   left-handed  and   right-handed  top
squarks~\cite{EWBAU}.

An  advantageous   feature  of  resonant  leptogenesis   is  that  the
predictions  for the BAU  are almost  independent of  any pre-existing
lepton- or baryon-number abundance.  This kind of fixed-line attractor
behaviour  is  a   consequence  of  the  quasi-in-thermal  equilibrium
dynamics governing the heavy Majorana neutrino sector. It results from
the fact that the heavy neutrino decay widths can be several orders of
magnitude  larger than  the expansion  rate  $H$ of  the Universe.   A
detailed analysis of this  dynamics was presented in \cite{PU2}, where
single   lepton-flavour   and   freeze-out  sphaleron   effects   were
systematically considered for the {\em first time}.  In particular, it
was  shown  that  single  lepton-flavour effects  resulting  from  the
Yukawa-neutrino couplings $h^{\nu}_{ij}$ can have a dramatic impact on
the predictions  for the  BAU, enhancing its  value by many  orders of
magnitude.  From the  model-building point of view, phenomenologically
rich  scenarios  are  now  possible  with  testable  implications  for
high-energy colliders~\cite{prodN} and low-energy observables, such as
$\mu  \to  e\gamma$, $\mu  \to  eee$ and  $\mu  \to  e$ conversion  in
nuclei~\cite{LFVN}.

We will not reiterate all  these results here, but only underline some
of the  key model-building aspects  related to the neutrino  sector of
the $F_D$-term  hybrid model.  The $F_D$-term hybrid  model contains a
$3\times 3$  Majorana mass matrix  $M_S$, which is SO(3)  symmetric at
the gauge-coupling  unification point  $M_X = M  \approx 10^{16}$~GeV,
i.e.\ $M_S  = m_N  {\bf 1}_3$.  The  parameter $m_N  = \rho v_S$  is a
universal Majorana  mass whose  natural value is  of the order  of the
soft  SUSY-breaking or  the electroweak  scale, i.e.~$m_N  \sim M_{\rm
SUSY}$ or $m_t$.   The SO(3) symmetry of the  heavy neutrino sector is
broken explicitly by the Yukawa neutrino couplings $h^{\nu}_{ij}$.  In
order to explain  the low-energy light neutrino data,  the breaking of
the  SO(3) symmetry should  proceed via  an intermediate  step, namely
SO(3) should  first break into  its subgroup SO(2)  $\simeq$ U(1)$_l$.
This can be achieved  by coupling all lepton doublets $L_{e,\mu,\tau}$
to  the  linear   combination:  $\frac{1}{\sqrt{2}}\,  (\nu_{2R}  +  i
\nu_{3R})$.   These  Yukawa  couplings   could  be  as  large  as  the
$\tau$-Yukawa coupling  $h_\tau$, i.e.~$h^{\nu}_{i2} =  i h^{\nu}_{i3}
\sim  10^{-2}$.   As  a  consequence  of the  U(1)$_l$  symmetry,  the
resulting   light  neutrino   mass  matrix   ${\bf   m}^\nu$  vanishes
identically  to  all orders  in  perturbation  theory.  The  remaining
U(1)$_l$ symmetry  can be  broken by smaller  Yukawa couplings  of the
order  of the  electron  Yukawa coupling  $h_e$, i.e.~$h^{\nu}_{i1}  =
\varepsilon_i \sim  10^{-6}$--$10^{-7}$, which arise  when one couples
$L_{e,\mu ,\tau}$ to $\nu_{1R}$~\cite{PUcomment}.

Further  breaking   of  the  U(1)$_l$  symmetry  is   induced  in  the
heavy-neutrino sector  by renormalization-group and  threshold effects
while running $M_S$ from $M$ to $m_t$~\cite{Branco}.  Thus, $M_S$ will
generically modify to:  $M_S = m_N {\bf 1}_3 +  \Delta M_S$, where one
typically has $(\Delta M_S)_{ij}/m_N \sim 10^{-5}$--$10^{-7}$.  Taking
the  effect of  U(1)$_l$-breaking parameters  $(\Delta  M_S)_{ij}$ and
$\varepsilon_i$ into account, one obtains a light neutrino mass matrix
which can comfortably accommodate  the low-energy light neutrino data,
e.g.~with an inverted hierarchical light neutrino spectrum~\cite{PU2}.
On the other hand, the  heavy neutrino sector of the $F_D$-term hybrid
model  consists  of  3  nearly  degenerate  heavy  Majorana  neutrinos
$N_{1,2,3}$ of  mass $m_{N_{1,2,3}} \approx m_N$, which  can give rise
to successful baryogenesis  through thermal electroweak-scale resonant
leptogenesis~\cite{PUcomment}.

\setcounter{equation}{0}
\section{Conclusions}\label{conclusions}

We have  studied $F$-term hybrid  inflation in a  novel supersymmetric
extension of the SM, to which  a subdominant FI $D$-term was added. We
called this particular form  of inflation $F_D$-term hybrid inflation.
The $F_D$-term hybrid model we  have been analyzing in this paper ties
the $\mu$-parameter  of the MSSM  to an SO(3) symmetric  Majorana mass
$m_N$, through the  VEV of the inflaton field.   As a consequence, the
model  predicts   {\em  naturally}  lepton-number   violation  at  the
electroweak scale.

In order to obtain predictions for the observables $P_{\cal R}^{1/2}$,
$n_s$    and   $\eta_B$    compatible    with   global    cosmological
analyses~\cite{Lyman},   as  well   as   interesting  particle-physics
phenomenology  that could  be  tested in  laboratory experiments,  one
needs to make  certain assumptions for the model  of $F_D$-term hybrid
inflation:
\begin{itemize}

\item[ (i)] Successful hybrid  inflation relies on the assumption that
  the inflaton field is displaced from its minimum in the beginning of
  inflation,  whereas all  other non-inflaton  fields have  zero VEVs,
  according to (\ref{initial}).

\item[ (ii)] The present $F_D$-term hybrid scenario utilizes a minimal
  K\"ahler  potential,  where  terms  of  order  $H^2  |S|^2$  in  the
  potential  are  set to  zero  or  assumed  to be  negligible.   This
  consideration introduces  some tuning  in general SUGRA  models with
  non-minimal K\"ahler potentials.

\item[(iii)] In order to get  a red-tilted spectrum with negative $n_s
  - 1$, one has to assume  that the radiative corrections dominate the
  slope of  the inflationary  potential.  This possibility  arises for
  superpotential  couplings:  $10^{-4} \stackrel{<}{{}_\sim}  \kappa,\
  \lambda,\ \rho \stackrel{<}{{}_\sim} 10^{-2}$.

\item[ (iv)] Even  though a bare $D$-tadpole may be  present as a bare
  parameter  in the  tree-level Lagrangian,  we have  considered here,
  however, the  possibility that such a term  is generated radiatively
  after  heavy degrees  of freedom  have been  integrated  out.  These
  heavy  degrees  of freedom  are  assumed  to  be Planck-mass  chiral
  superfields  which are  oppositely  charged under  the U(1)$_X$  and
  which break explicitly the  discrete charge symmetry discussed after
  (\ref{VFD}) and in Appendix~A.

\item[ (v)] We  have assumed that the coupling  $\rho$ of the inflaton
  to  neutrino superfields  is SO(3)  symmetric or  very close  to it.
  After  the inflaton  receives  a VEV,  one  ends up  with 3  nearly
  degenerate heavy  Majorana neutrinos with masses  at the electroweak
  scale. This enables  one to successfully address the  BAU within the
  thermal electroweak-scale  resonant leptogenesis framework  (see our
  discussion  in Section~\ref{BAU}).   As has  also been  discussed in
  Section~\ref{BAU}, if one assumes that the neutrino-Yukawa couplings
  $h^\nu_{ij}$ have a certain hierarchical structure controlled by the
  approximate  breaking of  global flavour  symmetries, the  model can
  have  further  testable  implications  for  $e^+e^-$  colliders  and
  low-energy experiments of lepton flavour and/or number violation.

\end{itemize}

The requirement for a sufficiently low reheat temperature $T_{\rm reh}
\stackrel{<}{{}_\sim} 10^9$~GeV, which does not lead to overproduction
of  gravitinos,   provides  an  important  constraint   on  the  basic
theoretical  parameters  $\kappa$, $\lambda$  and  $\rho$.  The  naive
limits on  these couplings derived from reheating  due to perturbative
inflaton  decay   are  very  strict,   i.e.~$\kappa,\  \lambda,\  \rho
\stackrel{<}{{}_\sim} 10^{-5}$. These limits may be completely avoided
by considering the late  decays of the U(1)$_X$ gauge-sector particles
which  are induced  by  a non-vanishing  FI  $D$-term $m^2_{\rm  FI}$.
Their  decay rates  depend  crucially on~$m^2_{\rm  FI}$. As  menioned
above  in  point~(iv)  and  in  Appendix~A, the  generation  of  a  FI
$D$-tadpole  and its  size may  be engineered  by  adding Planck-scale
heavy degrees  of freedom to the  theory and by  subjecting these into
extended $R$  symmetries.  In  this way, a  phase of  second reheating
takes place in the evolution of  the early Universe, which can lead to
a significant lowering of the reheat temperature even up to 1~TeV.

The  $F_D$-term  hybrid  model  with electroweak-scale  lepton  number
violation can easily be embedded  within a minimal SUGRA theory, where
all  soft  SUSY-breaking  parameters  are  constrained  at  the  gauge
coupling unification point $M_X$ which  can be chosen to be $M \approx
10^{16}$~GeV.  Instead,  electroweak baryogenesis  is not viable  in a
minimal  SUGRA  scenario  of  the  MSSM.  Moreover,  the  CP-odd  soft
SUSY-breaking phases required  for successful electroweak baryogenesis
face severe  constraints from the non-observation of  the electron and
neutron  electric  dipole  moments,   even  though  the  latter  arise
diagrammatically at the 2-loop level~\cite{CKP}.

The $F_D$-term hybrid model under discussion conserves $R$-parity. The
reason is that all  superpotential couplings either conserve the $B-L$
number  or  break it  by  even  number  of units.   Specifically,  the
operator   $\widehat{S}\widehat{N}_i\widehat{N}_i$  breaks  explicitly
$L$,  as  well  as  $B-L$,  by 2~units.   Consequently,  the  lightest
supersymmetric  particle  (LSP) of  the  spectrum  is  expected to  be
stable,  thus providing a  viable candidate  to address  the so-called
Cold Dark Matter  (CDM) problem.  The new aspect of  our model is that
right-handed   sneutrinos  could   be   the  LSPs,   opening  up   new
possibilities in the phenomenology of CDM and its detection.

From the  particle-physics point of  view and in the  low-energy limit
where  the  waterfall sector  has  decoupled  and the  $\rho$-coupling
neglected  for   simplicity,  the  $F_D$-term   hybrid  model  becomes
identical to the  so-called Minimal Nonminimal Supersymmetric Standard
Model (MNSSM)  in the decoupling  limit of a  large tadpole~\cite{PP}.
In   particular,    in   the    framework   of   WHI    discussed   in
Section~\ref{inflation},  the  coupling  $\lambda$  can  be  sizeable,
i.e.~$\lambda \sim 0.6$. In this  case, the Higgs phenomenology of the
MSSM will  modify drastically, despite  the decoupling of  the singlet
Higgs  states.  One  striking possibility  in  the MNSSM  is that  the
charged  Higgs boson  $H^+$ could  be lighter  than the  SM-like Higgs
boson~\cite{PP2},    thus     pointing    to    particular    collider
phenomenologies~\cite{DP}.   However,   even  within  the  traditional
scenario   of  CHI,   where  $\kappa,   \lambda  \stackrel{<}{{}_\sim}
10^{-2}$, the $F_D$-term hybrid model will favour particular benchmark
scenarios  of the  MSSM.  For  example, if  $\lambda \gg  \kappa$, the
$F_D$-term hybrid model may account  for a possible large value of the
$\mu$-parameter.   Specifically, if  $\lambda  = 4  \kappa$, one  gets
from~(\ref{Send}) the hierarchy $\mu \approx 4 M_{\rm SUSY}$, which is
the so-called CPX  benchmark scenario~\cite{CPX} describing maximal CP
violation in the MSSM Higgs sector at low and moderate values of $\tan
\beta$.

A  possible  natural  solution  to the  famous  cosmological  constant
problem  is  expected to  provide  further  constraints  on the  model
building of cosmologically viable models in future.  Nevertheless, the
$F_D$-term hybrid  model presented in  this paper constitutes  a first
attempt  towards the  formulation  of a  minimal Particle-Physics  and
Cosmology  Standard  Model, whose  validity  could,  in principle,  be
tested   in   laboratory  experiments   and   further  vindicated   by
astronomical observations.

\bigskip\bigskip

\subsection*{Acknowledgements}

We  thank Arjun  Berera, Rudnei  Ramos and  Antonio Riotto  for useful
discussions. We also thank Constantine Pallis for collaboration in the
early stages of this project.  AP~dedicates this work to the memory of
Darwin  Chang, an invaluable  friend and  collaborator.  This  work is
supported in part by the PPARC research grants: PPA/G/O/2002/00471 and
PP/C504286/1.

\newpage

\def\theequation{\Alph{section}.\arabic{equation}}
\begin{appendix}

\setcounter{equation}{0}
\section{{\boldmath $D$}--Term Engineering}\label{Dappendix}

The generation and the size of  a $D$-term may be engineered by adding
Planck-scale heavy degrees of freedom  to the theory and by subjecting
these into extended $R$ symmetries.

To elucidate our  point, let us first consider a  model augmented by a
pair of oppositely charged superfields $\widehat{\overline{X}}_{1,2}$,
with    U(1)$_X$    charges:    $Q(\widehat{\overline{X}}_2)    =    -
Q(\widehat{\overline{X}}_1 ) =  Q(\widehat{X}_1) = - Q(\widehat{X}_2 )
= 1$. The extended superpotential $W$ of our interest is
\begin{equation}
  \label{Wdterm}
 W \ =\ \kappa\, \widehat{S}\, \Big( \widehat{X}_1
\widehat{X}_2\:  -\: M^2\Big)\ +\ \xi\, m_{\rm Pl}\,
\widehat{\overline{X}}_1\,\widehat{\overline{X}}_2\ +\
\xi_1\, \frac{ ( \widehat{\overline{X}}_1\widehat{X}_1 )^2}{2\, m_{\rm
    Pl}}\ +\ \xi'_1\, 
   \frac{ ( \widehat{\overline{X}}_2\widehat{X}_2 )^2}{2\, m_{\rm Pl}}\ .
\end{equation}
This form of  the superpotential may be enforced  by the $R$ symmetry:
$\widehat{S}  \to e^{i\alpha}\,  \widehat{S}$,  $\widehat{X}_{1,2} \to
e^{\pm  i\beta}\,\widehat{X}_{1,2}$, $\widehat{\overline{X}}_{1,2} \to
e^{i(\frac{a}{2}    \mp    \beta)}\,    \widehat{\overline{X}}_{1,2}$,
$\widehat{L}   \to   e^{i\alpha}\,   \widehat{L}$,  $\widehat{Q}   \to
e^{i\alpha}\, \widehat{Q}$,  with $W  \to e^{i\alpha} W$.   As before,
all  remaining fields  are  considered  to be  neutral  under the  $R$
symmetry.  Notice  that the same $R$-symmetry allows  for the operator
$\kappa'  S  (\widehat{X}_1   \widehat{X}_2  )^2/m^2_{\rm  Pl}$.   The
presence  of  this  superpotential  term can  trigger  shifted  hybrid
inflation,  where the  gauge  symmetry U(1)$_X$  is  broken along  the
inflationary trajectory,  thereby inflating away  unwanted topological
defects~\cite{JKLS}.

A  $D$-term   will  now  be   generated  after  integrating   out  the
Planck-scale     superfields    $\widehat{\overline{X}}_{1,2}$.    The
loop-induced $D$-tadpole $m^2_{\rm FI}$ is found to be
\begin{equation}
  \label{FIdterm}
m^2_{\rm FI}\ \approx\ \frac{\xi^2_1 - \xi'^2_1}{8\pi^2}\
\frac{M^4}{m^2_{\rm Pl}}\ \ln\left(\frac{m_{\rm Pl}}{M}\right)\ .
\end{equation}
For    $M   =    10^{16}$~GeV,    we   find    that   $m_{\rm    FI}/M
\stackrel{<}{{}_\sim}      10^{-3}$,      for     $\xi_1,\      \xi'_1
\stackrel{<}{{}_\sim}  0.3$. Observe  that  if $\xi_1  = \xi'_1$,  the
discrete  charge symmetry  discussed after  (\ref{VFD})  gets restored
again and $m_{\rm FI}$ vanishes identically.

The size of the $D$-term may be suppressed further, if the Planck-mass
chiral   superfields  $\widehat{\overline{X}}_{1,2}$   possess  higher
U(1)$_X$  charges.   In general,  one  may  assume  that the  U(1)$_X$
charges         of         $\widehat{\overline{X}}_{1,2}$         are:
$Q(\widehat{\overline{X}}_2)  = -  Q(\widehat{\overline{X}}_1 )  = n$,
where     $n\ge     1$.     In     addition,     we    require     for
$\widehat{\overline{X}}_{1,2}$ to transform under U(1)$_R$ as follows:
\begin{equation}
  \label{Rsymn}
\widehat{\overline{X}}_{1,2}\ \to\ e^{\frac{i}2\, [a \, \mp\, (n+1)
\beta ]}\; \widehat{\overline{X}}_{1,2}\; ,
\end{equation}
while   $\widehat{S}$,  $\widehat{X}_{1,2}$   and  all   other  fields
transform   as   before.   With   this   symmetry   restriction,   the
superpotential reads:
\begin{equation}
  \label{Wdtermn}
W \ =\ \kappa\, \widehat{S}\, \Big( \widehat{X}_1 \widehat{X}_2\: -\:
M^2\Big)\ +\ \xi\, m_{\rm Pl}\,
\widehat{\overline{X}}_1\,\widehat{\overline{X}}_2\ +\ \xi_n\, \frac{
(\widehat{\overline{X}}_1)^2\, (\widehat{X}_1)^{n+1}}{2\,m^n_{\rm Pl}}\ +\
\xi'_n\, \frac{ (
  \widehat{\overline{X}}_2)^2\,(\widehat{X}_2)^{n+1}}{2\,m^n_{\rm Pl}}\ . 
\end{equation}
In this case, the loop-induced $D$-term is given by 
\begin{equation}
  \label{FIdtermn}
m^2_{\rm FI}\ \approx\ \frac{\xi^2_n - \xi'^2_n }{8\pi^2}\
\frac{M^{2(n+1)}}{m^{2n}_{\rm Pl}}\ \ln\left(\frac{m_{\rm Pl}}{M}\right)\ .
\end{equation}
To obtain  a small  ratio $m_{\rm FI}/M  \sim 10^{-6}$,  with $\xi_n,\
\xi'_n \sim 1$, one would need  $n = 5,\ 6$.  Finally, it is important
to remark that the loop-induced  $D$-term does not lead to spontaneous
breakdown of global supersymmetry.

\end{appendix}

\end{document}